\documentclass[10pt,pdftex]{iopart}
\usepackage{iopams}

\usepackage{tensind,graphicx}
\usepackage{gensymb,bbm,wasysym,amssymb}
\usepackage{mathptmx}      % use Times fonts if available on your TeX system
\usepackage{color}
\usepackage{url}
\usepackage{setstack}
\usepackage{units}

\newcounter{theExercise} 
\newenvironment{cpExercise}{\noindent\begin{minipage}[t]{4.5in}\textbf{Exercise \arabic{theExercise}:}\it \stepcounter{theExercise}}{\end{minipage}\vspace*{0.2cm}}
\newenvironment{cpResult}{\begin{minipage}[t]{4.5in}\textbf{Result:}}{\end{minipage}\vspace*{0.5cm}}

\def\pdiff#1#2{\frac{\partial #1}{\partial #2}}

\def\vecThree#1#2#3{\left(\!\begin{array}{c} #1\\ #2\\ #3\end{array}\!\right)}

\begin{document}
\title{Charged particles constrained to a curved surface}

\author{Thomas M{\"u}ller}
\address{
  Visualisierungsinstitut der Universit\"at Stuttgart (VISUS)\\
  Allmandring 19, 70569 Stuttgart, Germany
}
\ead{Thomas.Mueller@visus.uni-stuttgart.de}

\author{J{\"o}rg Frauendiener}
\address{
  Department of Mathematics \& Statistics, University of Otago,\\
  P.O. Box 56, Dunedin 9010, New Zealand
}
\ead{joergf@maths.otago.ac.nz}

% -----------------------------------------------------------------
%                            Abstract
% -----------------------------------------------------------------
\begin{abstract}
We study the motion of charged particles constrained to arbitrary two-dimensional curved surfaces but interacting in three-dimensional space via the Coulomb potential. To speed-up the interaction calculations, we use the parallel compute capability of the Compute Unified Device Architecture (CUDA) of todays graphics boards. The particles and the curved surfaces are shown using the Open Graphics Library (OpenGL). The paper is intended to give graduate students, who have basic experiences with electrostatics and the Lagrangian formalism, a deeper understanding in charged particle interactions and a short introduction how to handle a many particle system using parallel computing on a single home computer.
\end{abstract}

% -----------------------------------------------------------------
%                            PACS
%
%   http://publish.aps.org/PACS
% -----------------------------------------------------------------
\pacs{01.50.hv,02.40.Hw,02.60.Cb,45.50.-j}

%01.50.hv 	Computer software and software reviews
%02.40.Hw 	Classical differential geometry 
%02.60.Cb 	Numerical simulation; solution of equations 
%45.50.-j 	Dynamics and kinematics of a particle and a system of particles
%45.20.Jj 	Lagrangian and Hamiltonian mechanics 

%\keywords{charged particles}
%\maketitle

\submitto{EJP}

% -----------------------------------------------------------------
%   Introduction
% -----------------------------------------------------------------
\section{Introduction}\label{sec:intro}
The original idea of the Thomson problem\cite{thomson,Ashby1986} of 1904 was to find equilibrium positions of $N$ charges constrained to a spherical surface interacting with Coulomb's law. 
More than 100 years later there is still interest in finding minimum energy configurations of such assemblies, see e.g. \cite{Whyte1952,Marx1970,Erber1991,Altschuler2005,Altschuler2006,Backofen2010,Lakhbab2012}.
There is also an interactive Java applet by Bowick et al.~\cite{thomsonApplet} from Syracuse university (NY) to find minimum configurations for the more general $r^{-n}$ potential using several different minimization algorithms.

In this paper, we generalize the Thomson problem to arbitrary curved non-self-penetrating parametrized two-dimensional surfaces that are embedded in three-dimensional Cartesian space. We present the mathematical details to study the motion and minimum energy configurations of an arbitrary number of charged particles constrained to these surfaces, and we briefly describe how to implement the resulting N-body simulation using the Compute Unified Device Architecture (CUDA) of todays graphics boards. The generalization to $r^{-n}$ potentials is left as exercise for motivated students having basic experiences with electrostatics and the Lagrangian formalism. 

The structure of the paper is as follows. In Sec.~\ref{sec:trajectories} we briefly discuss the details for the parametrization of trajectories on curved surfaces. The Lagrangian for $N$ charged particles on this surface interacting via Coulomb is compiled in Sec.~\ref{sec:lagrangian}. From this Lagrangian, we derive in section \ref{sec:eom} the equations of motion for each particle. Additionally, we extend the equation of motion when there is also an external electric and magnetic field. In Sec.~\ref{sec:simvis} we discuss some implementation details for the N-body simulation and the subsequent visualization. Several examples and some feasible exercises are presented in sections~\ref{sec:examples} and ~\ref{sec:exercises}.

The source code to reproduce the examples in this paper is written in C/C++/CUDA and is freely available from (\url{http://www.vis.uni-stuttgart.de/chapacs}). It can be compiled on Linux and Windows systems.

% -----------------------------------------------------------------
%   Particle constrained to a curved surfaces
% -----------------------------------------------------------------
\section{Particle constrained to a curved surface}\label{sec:trajectories}
\begin{figure}[ht]
  %\centering\includegraphics[scale=0.8]{pics/parametrization}
  \centering\includegraphics[scale=0.8]{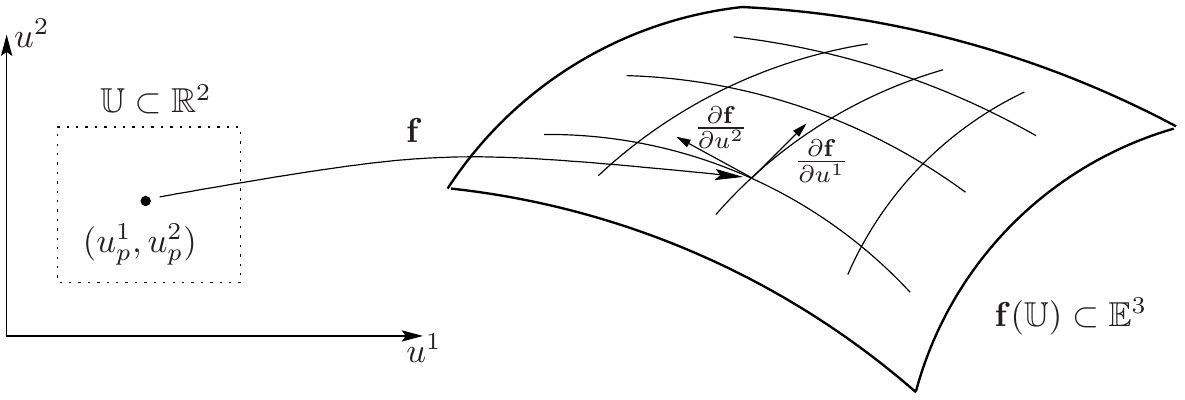}
  \caption{Parametrized surface $S\subset\mathbb{E}^3$.}
  \label{fig:paramCurve}
\end{figure}

The trajectory $\gamma$ of a particle that is constrained to a two-dimensional curved surface $S$ can be derived from the Lagrangian formalism using generalized coordinates that are adapted to the surface. These adapted coordinates are usually the parameters $(u^1,u^2)$ that are used as surface parametrization 
\begin{equation}
  \mathbf{f}: \mathbb{U}\rightarrow\mathbb{E}^3,\quad (u^1,u^2)\mapsto\mathbf{f}\left(u^1,u^2\right)\in S\subset\mathbb{E}^3,
  \label{eq:surfFunction}
\end{equation}
where $\mathbb{U}$ is some open domain in $\mathbb{R}^2$ and $\mathbb{E}^3$ is the three-dimensional Euclidean space, see figure~\ref{fig:paramCurve}.

Instead of using the Cartesian coordinates $\mathbf{x}(t)$ for the particle's trajectory, we can now describe $\gamma$ also by means of a curve $\vec{u}(t)$ in $\mathbb{U}$, 
\begin{equation}
   \gamma:\quad t\mapsto \mathbf{x}(t) = \mathbf{f}(\vec{u}(t))=\mathbf{f}\left(u^1(t),u^2(t)\right)
\end{equation}
where $t\in[t_i,t_f]$ denotes time. The velocity $\mathbf{v}(t)$ of the particle is then obtained by
\begin{equation}
  \mathbf{v}(t) = \mathbf{\dot{x}}(t) = \frac{d}{dt}\mathbf{x}(t) = \frac{d}{dt}\mathbf{f}\left(\vec{u}(t)\right) = \sum_{i=1}^2\pdiff{\mathbf{f}}{u^i}\frac{du^i}{dt} = \sum_{i=1}^2\pdiff{\mathbf{f}}{u^i}\dot{u}^i.
  \label{eq:timeDiffF}
\end{equation}

% -----------------------------------------------------------------
%   Particle constrained to a curved surfaces
% -----------------------------------------------------------------
\section{Lagrangian for $N$ charged particles on a curved surface}\label{sec:lagrangian}
The Lagrangian for $N$ charged particles that are constrained to a two-dimensional curved surface can be obtained in the usual way,
\begin{equation}
  L = T - V,
\end{equation}
where $T$ is the total kinetic energy and $V$ is the total interaction energy. For $N$ particles the kinetic energy is
\begin{equation}
  T = \frac{1}{2}\sum_{A=1}^N M_A\left<\mathbf{\dot{x}}_A,\mathbf{\dot{x}}_A\right>,
  \label{eq:kinEnergy}
\end{equation}
where $M_A$ denotes the mass of particle $A$ and $\left<\cdot,\cdot\right>$ is the standard scalar product in $\mathbb{E}^3$. With \eref{eq:timeDiffF} we can write equation \eref{eq:kinEnergy} as
\begin{equation}
\fl T = \frac{1}{2}\sum_{A=1}^N M_A\sum_{i,k=1}^2\left<\pdiff{\mathbf{f}}{u^i}(\vec{u}_A),\pdiff{\mathbf{f}}{u^k}(\vec{u}_A)\right>\dot{u}_A^i\dot{u}_A^k = \frac{1}{2}\sum_{A=1}^N M_A\sum_{i,k=1}^2g_{ik}(\vec{u}_A)\dot{u}_A^i\dot{u}_A^k.
\end{equation}
Here, we have introduced the functions 
\begin{equation}
   g_{ik}(\vec{u})=\left<\pdiff{\mathbf{f}}{u^i}(\vec{u}),\pdiff{\mathbf{f}}{u^k}(\vec{u})\right>
   \label{eq:metricCoeffs}
\end{equation}
on $\mathbb{U}$. They can be interpreted as the coefficients of the metric induced on $S$ expressed in the coordinate system $(u^1,u^2)$ (see e.g. \cite{doCarmo}).

The interaction energy is obtained as
\begin{equation}
  V = \sum_{\underset{A<B}{A,B=1}}^N V_{AB},
  \label{eq:fieldEnergy}
\end{equation}
where $V_{AB}$ is the interaction energy between particles $A$ and $B$. In the present case we are only concerned with the electrostatic (Coulomb) interaction,
\begin{equation}
  V_{AB} = \frac{1}{4\pi\epsilon_0}\frac{Q_AQ_B}{\|\mathbf{x}_A-\mathbf{x}_B\|} = \frac{1}{4\pi\epsilon_0}\frac{Q_AQ_B}{\|\mathbf{f}(\vec{u}_A)-\mathbf{f}(\vec{u}_B)\|}
\end{equation}
with $Q_A$ and $Q_B$ being the charges of particle $A$ or $B$, respectively.
Since $V_{AB}=V_{BA}$ and since we ignore the self-energy of a particle with itself, i.e. $V_{AA}=0$, we can write
\begin{equation}
  V = \frac{1}{2}\sum_{\underset{A\neq B}{A,B=1}}^N\frac{1}{4\pi\epsilon_0}\frac{Q_AQ_B}{\|\mathbf{f}(\vec{u}_A)-\mathbf{f}(\vec{u}_B)\|}.
\end{equation}
Thus, the total Lagrangian reads
\begin{equation}
L = \frac{1}{2}\sum_{A=1}^N M_A\sum_{i,k=1}^2g_{ik}(\vec{u}_A)\dot{u}_A^i\dot{u}_A^k - \frac{1}{2}\sum_{\underset{A\neq B}{A,B=1}}^N\frac{1}{4\pi\epsilon_0}\frac{Q_AQ_B}{\|\mathbf{f}(\vec{u}_A)-\mathbf{f}(\vec{u}_B)\|}.
\end{equation}

% -----------------------------------------------------------------
%     Equations of motion 
% -----------------------------------------------------------------
\section{Equations of motion}\label{sec:eom}
We obtain the equations of motion for particle $A$ by evaluating the Euler-Lagrange equations
\begin{equation}
 0 = \frac{d}{dt}\pdiff{L}{\dot{u}_A^i}-\pdiff{L}{u_A^i}\quad\mbox{for}\quad i=\{1,2\}.
 \label{eq:eulerLagrange}
\end{equation}
We compute the two terms separately. First, the partial derivative of $L$ with respect to $\dot{u}_A^i$ is
\begin{equation}
  \pdiff{L}{\dot{u}_A^i} = M_A\sum_{k=1}^2g_{ik}(\vec{u}_A)\dot{u}_A^k
\end{equation}
and its time derivative reads
\begin{equation}
  \frac{d}{dt}\pdiff{L}{\dot{u}_A^i}=M_A\left(\sum_{l,k=1}^2\pdiff{g_{ik}}{u^l}(\vec{u}_A)\dot{u}_A^l\dot{u}_A^k + \sum_{k=1}^2g_{ik}(\vec{u}_A)\ddot{u}_A^k\right).
  \label{eq:el1}
\end{equation}
Next, to compute the partial derivative of $L$ with respect to $u_A^i$ we need the partial derivative $\partial r^{-1}/\partial u_A^i$ with $r:=\|\mathbf{f}_A-\mathbf{f}_B\|=\|\mathbf{f}(\vec{u}_A)-\mathbf{f}(\vec{u}_B)\|=\sqrt{\left<\mathbf{f}(\vec{u}_A)-\mathbf{f}(\vec{u}_B),\mathbf{f}(\vec{u}_A)-\mathbf{f}(\vec{u}_B)\right>}$,
\begin{eqnarray}
   \pdiff{}{u_A^i}r^{-1} = -\frac{1}{2}r^{-3}\pdiff{}{u_A^i}\left<\mathbf{f}_A-\mathbf{f}_B,\mathbf{f}_A-\mathbf{f}_B\right> = -r^{-3}\left<\mathbf{f}_A-\mathbf{f}_B,\pdiff{\mathbf{f}_A}{u_A^i}\right>.
\end{eqnarray}
Hence,
\begin{eqnarray}
  \pdiff{L}{u_A^i} = \frac{1}{2}M_A\sum_{l,k=1}^2\pdiff{g_{lk}}{u^i}(\vec{u}_A)\dot{u}_A^l\dot{u}_A^k + \frac{Q_A}{4\pi\epsilon_0}\sum_{\underset{B\neq A}{B=1}}^N\frac{Q_B}{r^3}\left<\mathbf{f}_A-\mathbf{f}_B,\pdiff{\mathbf{f}_A}{u_A^i}\right>.
  \label{eq:el2}
\end{eqnarray}
Bringing equations \eref{eq:el1} and \eref{eq:el2} together and solving for $\ddot{u}_A^i$ we obtain
\begin{eqnarray}
 \fl \nonumber \ddot{u}_A^i &= -\frac{1}{2}\sum_{j,k,l=1}^2g^{ij}(\vec{u}_A)\left(2\pdiff{g_{jk}}{u^l}(\vec{u}_A)-\pdiff{g_{lk}}{u^j}(\vec{u}_A)\right)\dot{u}_A^l\dot{u}_A^k\\
 \fl  &\quad +\frac{Q_A}{4\pi\epsilon_0M_A}\sum_{\underset{B\neq A}{B=1}}^N\frac{Q_B}{\|\mathbf{f}(\vec{u}_A)-\mathbf{f}(\vec{u}_B)\|^3}\left<\mathbf{f}(\vec{u}_A)-\mathbf{f}(\vec{u}_B),\sum_{j=1}^2g^{ij}\pdiff{\mathbf{f}}{u^j}(\vec{u}_A)\right>.
   \label{eq:eom}
\end{eqnarray}
Here, $g^{ij}$ are the matrix elements of the inverse of the matrix $g_{jk}$, so that we have the identity $\sum_{j=1}^2g^{ij}g_{jk}=\delta_k^i$ with $\delta_k^i$ being the Kronecker-$\delta$. The combination of derivatives of the metric at the beginning of \eref{eq:eom} is usually abbreviated by
\begin{equation}
  \frac{1}{2}\sum_{j,k,l=1}^2g^{ij}\left(2\pdiff{g_{jk}}{u^l}-\pdiff{g_{lk}}{u^j}\right)\dot{u}^l\dot{u}^k = \sum_{k,l=1}^2\Gamma_{lk}^i\dot{u}^l\dot{u}^k.
\end{equation}
The Christoffel symbols $\Gamma_{lk}^i$ have the geometric meaning of defining the notion of parallel displacement on the surface $S$. Note, that the functions $g_{ij}$, $g^{ij}$, and $\Gamma_{lk}^i$ can all be precomputed, once the parametrization $\mathbf{f}$ is given. They are intrinsic properties of the surface $S$. \ref{appsec:surfExp} lists the metric coefficients and Christoffel symbols for the sphere and the torus which we use in section \ref{sec:examples}.

If there is also an external electric or magnetic field, the interaction energy $V$ must be extended to
\begin{equation}
 V = \sum_{\underset{A<B}{A,B=1}}^N V_{AB} + \sum_{A=1}^N\left[Q_A\phi(\mathbf{f}(\vec{u}_A),t) - Q_A\left<\mathbf{\dot{f}}(\vec{u}_A),\mathbf{A}(\mathbf{f}(\vec{u}_A),t)\right>\right]
 \label{eq:lagrangian}
\end{equation}
with the electric potential $\phi$, and the magnetic vector potential $\mathbf{A}$. These potentials are related to their field values via $\mathbf{E}=-\mathbf{\nabla}\phi-\partial\mathbf{A}/\partial t$ and $\mathbf{B}=\mathbf{\nabla}\times\mathbf{A}$.
The final equations of motion for particle $A$ then reads
\begin{eqnarray}
  \fl \nonumber \ddot{u}_A^i &= -\sum_{k,l=1}^2\Gamma_{lk}^i\dot{u}_A^l\dot{u}_A^k +\frac{Q_A}{4\pi\epsilon_0M_A}\sum_{\underset{B\neq A}{B=1}}^N\frac{Q_B}{\|\mathbf{f}(\vec{u}_A)-\mathbf{f}(\vec{u}_B)\|^3}\left<\mathbf{f}(\vec{u}_A)-\mathbf{f}(\vec{u}_B),\sum_{j=1}^2g^{ij}\pdiff{\mathbf{f}}{u^j}(\vec{u}_A)\right>\\
   \fl &\quad + \frac{Q_A}{M_A}\left<\mathbf{E}+\mathbf{\dot{f}}(\vec{u}_A)\times\mathbf{B},\sum_{j=1}^2g^{ij}\pdiff{\mathbf{f}}{u^j}(\vec{u}_A)\right>.
   \label{eq:eomAll}
\end{eqnarray}

Here, we neglect that the charged particle motion itself yields a magnetic field that could influence the motion of the other particles. Furthermore, we do not take account of energy loss due to electromagnetic radiation caused by accelerated motion of the charged particles, but we add an artificial frictional term, see \ref{appsec:partFriction}.

% -----------------------------------------------------------------
%    
% -----------------------------------------------------------------
\section{N-body simulation and visualization}\label{sec:simvis}
As long as the number of particles in an N-body simulation is in the order of a few thousands, we do not need any specific acceleration algorithm and/or approximation procedure, but we can calculate the N-body interaction by brute force: every particle interacts with every other particle. The computation time, however, increases quadratically with the number of particles. A first step to accelerate the computation is to handle each particle by a separate compute unit that has to integrate the equation of motion \eref{eq:eomAll} for this particle. The only prerequisite is that all compute units must have access to all particle positions and velocities which can be achieved using a shared memory system. 
Today, virtually all standard home computers and even high-end smartphones have at least a dual core processor inside that have access to shared memory. 
Parallelization of computation can then be realized, for example, by the programming interface \emph{OpenMP}\cite{openmp} that splits the computation into several threads.

Much higher parallelization can be achieved using the compute capability of modern graphics hardware. The \emph{Compute Unified Device Architecture} (CUDA) or the \emph{Open Compute Language} (OpenCL) offer a C-like programming interface in order to use the graphics processing units (GPUs) for general purpose computations.
Even without sophisticated algorithms for an efficient memory access, GPU computation leads to an enormous speed-up. Together with the Open Graphics Library (OpenGL) we can explore physical simulations at interactive frame rates.

\begin{figure}[ht]
  %\centering\includegraphics[scale=0.7]{pics/structure}
  \centering\includegraphics[scale=0.7]{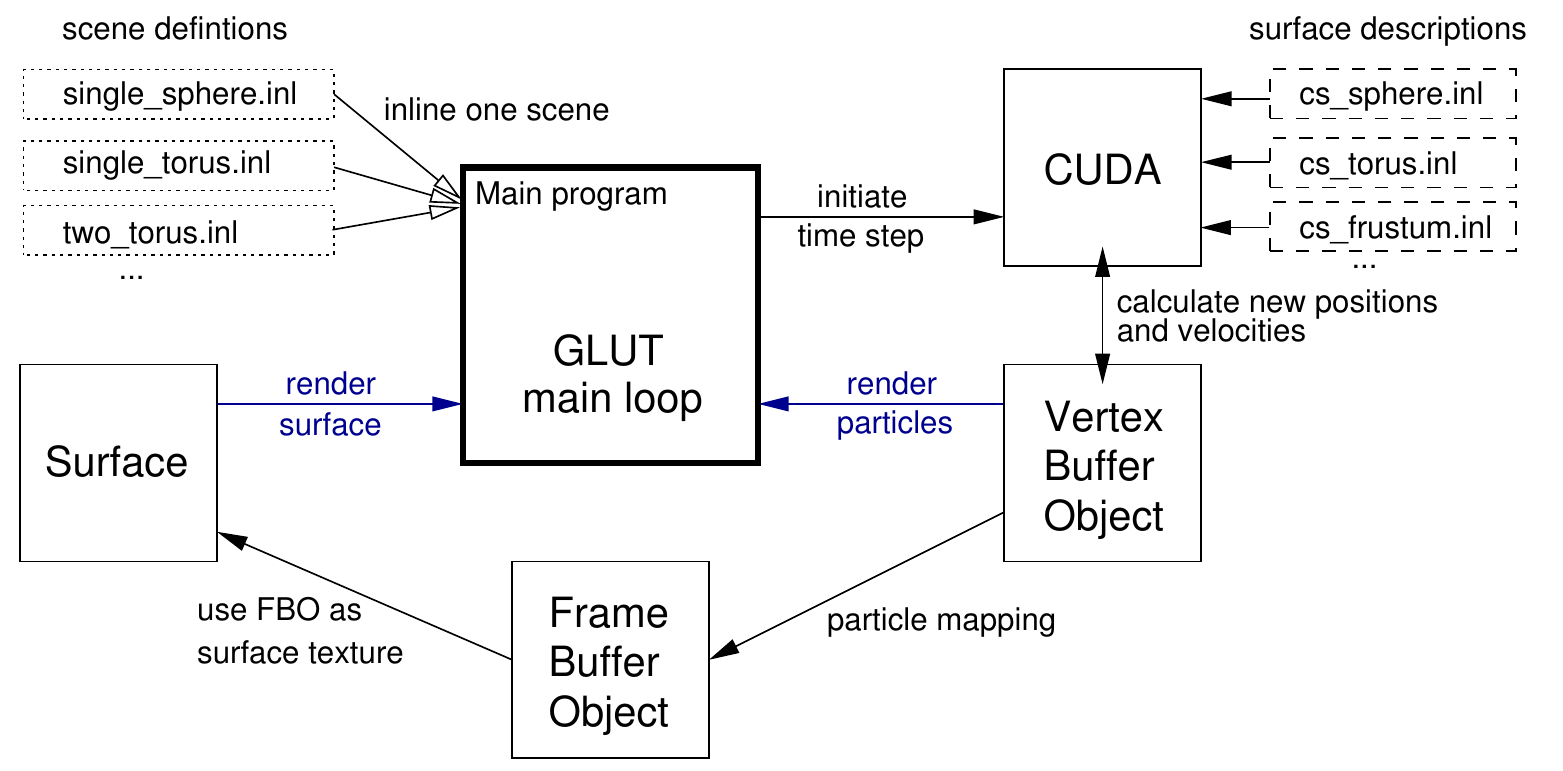}
  \caption{Basic structure of the program.}
  \label{fig:progStructure}
\end{figure}

The basic structure of our implementation is shown in figure~\ref{fig:progStructure}. The basic block is the GLUT~\cite{glut} main loop which acts on key strokes and mouse events, and which initiates rendering new frames. 
The ``CUDA'' block includes all surface descriptions $(\mathbf{f},\partial\mathbf{f},g_{ij},\Gamma_{ij}^k)$ and calculates a single time step using either a standard Runge-Kutta second or fourth order method, see e.g. Press et al.~\cite{Press1992}. To calculate a time step, it uses the particle positions $u^k$ and velocities $\dot{u}^k$ stored within a \emph{Vertex Buffer Object (VBO)} which can be directly accessed by CUDA and OpenGL. The new particle positions can then be rendered directly or they can be mapped onto their corresponding surfaces. This mapping is realized by means of a \emph{Frame Buffer Object} (FBO). This FBO is an internal rectangular image (display) which, in our case, represents the domain $\mathbb{U}$. This image is then used to texturize the surface. 

\begin{figure}[ht]
  \includegraphics[scale=0.3]{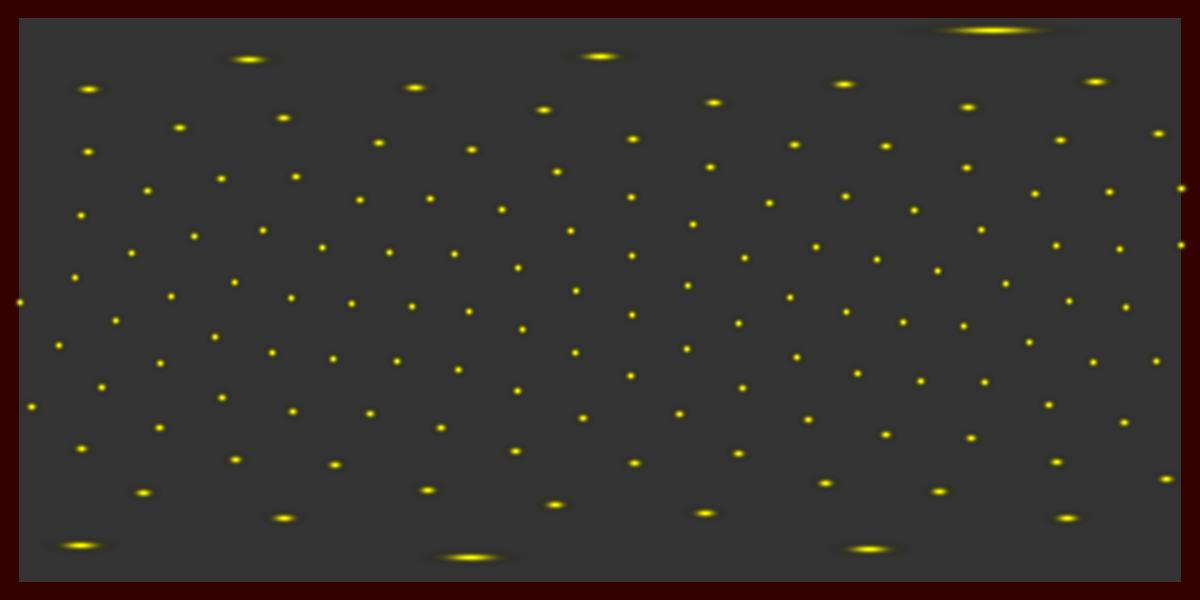}
  \caption{Particle rendering as Gaussian splats with radii $d\vartheta = ds/r$ and $d\varphi = ds/(r\sin\vartheta)$ for fixed value $ds$. This FBO image can be mapped onto a sphere, see figure~\ref{fig:splatArtefacts}. The inner gray rectangle covers the whole domain of the sphere $u^1=\varphi\in[0,2\pi)$, $u^2=\vartheta\in(0,\pi)$. The dark red border (color online) prevents particle splats near the domain's boundary from being clipped.}
  \label{fig:sphereTex}
\end{figure}

\begin{figure}[ht]
  \includegraphics[scale=0.2]{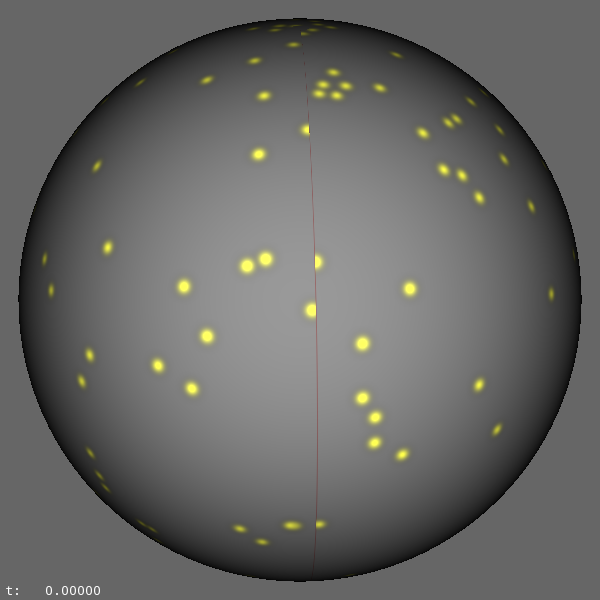}
  \includegraphics[scale=0.2]{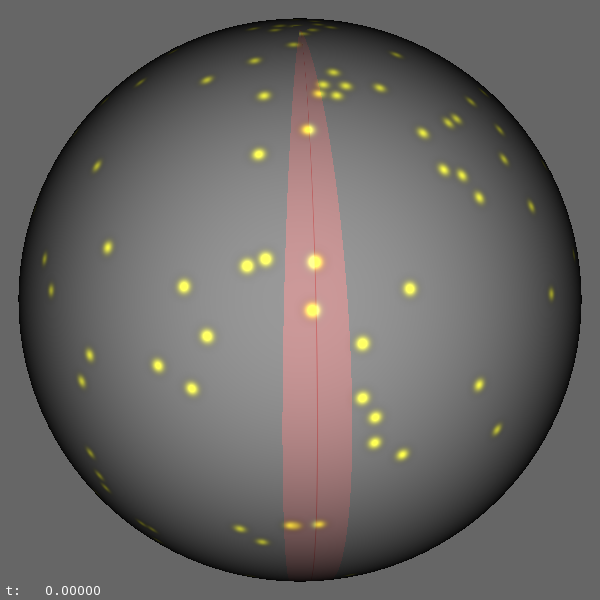}
  \includegraphics[scale=0.2]{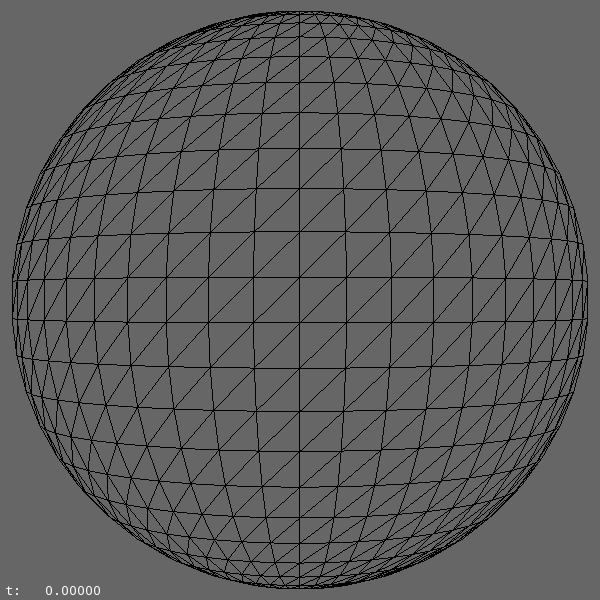}
  \caption{Particle rendering as Gaussian splats without (left) and with (center) extended domain, compare figure~\ref{fig:sphereTex}. Without extended domain the splats are clipped. Right: wireframe view of the sphere.}
  \label{fig:splatArtefacts}
\end{figure}

As an example, figure~\ref{fig:sphereTex} shows particles (yellow splats) projected onto a rectangular image (FBO) that is used as texture for the sphere. The inner gray rectangle covers the whole domain of the sphere $u^1=\varphi\in[0,2\pi)$, $u^2=\vartheta\in(0,\pi)$. The dark red border slightly expands the domain to prevent particle splats near the domain's boundary from being clipped, see figure~\ref{fig:splatArtefacts}.
The splats have to be distorted by means of the inverse metric of the surface to let them appear as circular splats when mapped onto the surface.

The ``Surface'' block is responsible for drawing the surfaces themself. For that, we uniformly sample the domain using quads. These quads are then split into two triangles which are transformed by means of the surface function $\mathbf{f}$ of \eref{eq:surfFunction} within a so called vertex shader (see OpenGL Shading Language~\cite{opengl}). Figure~\ref{fig:splatArtefacts}~(right) shows the resulting wireframe of a sphere.

To keep the code simple, we do not have any sophisticated scene description language but implement each scene in a separate ``.inl''-file. A specific scene and its particular scene parameters have to be chosen at compile time. Each ``.inl''-file must have three functions: {\tt init\_Objects()} defines all surfaces and assigns IDs to them; {\tt set\_Particles()} registers the number of particles, their parameters (mass, charge, initial position, initial velocity), and the ID of the surface they belong to; {\tt set\_Supplement()} offers the possibility to change the camera parameters or the size of the window. The particle data could also be loaded from file when starting the program.

%Center of charge
%\begin{equation}
  %\vec{r}_c = \sum_{i=1}^Nq_i\vec{f}_i \bigg/ \sum_{i=1}^Nq_i
%\end{equation}

% -----------------------------------------------------------------
%
% -----------------------------------------------------------------
\section{Examples}\label{sec:examples}
In the following examples, we use the explicit second-order Runge-Kutta method without step size control and double precision floating-point numbers to integrate the equation of motion \eref{eq:eomAll}. The units in use are explained in \ref{appsec:units}. The divergence problem of close encounters due to the $1/\|\mathbf{f}_A-\mathbf{f}_B\|$ term is handled by modifying the Coulomb potential:
\begin{equation}
  \tilde{V}_{AB} = \frac{1}{4\pi\epsilon_0}\frac{Q_AQ_B}{\sqrt{\|\mathbf{f}_A-\mathbf{f}_B\|^2+\varepsilon^2}}\quad\mbox{with}\quad \varepsilon\ll 1.
\end{equation}
In principle, we should also take into account the magnetic field $\vec{B}$ generated by a moving charged particle which follows from the Biot-Savart law
\begin{equation}
  \vec{B} = \frac{\mu_0}{4\pi}\frac{Q_A\vec{v}_A\times\vec{r}}{r^3},
\end{equation}
where $\vec{r}$ is the relative distance of an other particle to particle $A$'s current position.
Then, from Lorentz force equation, the resulting acceleration of a particle $B$ with mass $M_B$ and current velocity $\vec{v}_B$ would be
\begin{equation}
  \vec{a} = \frac{Q_B}{M_B}\vec{v}_B\times\vec{B} = \frac{\mu_0e^2}{4\pi m_e}\frac{q_Aq_B}{m_B}\frac{\vec{v}_B\times(\vec{v}_A\times\vec{r})}{r^3},
  \label{eq:lorentz}
\end{equation}
where $\mu_0e^2/(4\pi m_e)\approx 2.81806\cdot 10^{-15}\unit{m}$. As we will see, this acceleration can be neglected.

\subsection{Particles on a sphere}
As a first example, we consider a Thomson problem situation where $N=128$ charged particles (electrons) are located on a sphere. For numerical reasons, we set the radius $r$ of the sphere such that the mutual acceleration $a$ between two particles due to Coulomb,
\begin{equation}
  a = \frac{e^2}{4\pi\epsilon_0m_e}\frac{1}{d^2}=\kappa\frac{1}{d^2},
\end{equation}
is roughly in the same order of magnitude as their distance $d\approx r$. For a sphere of radius $r=1\unit{m}$, this acceleration is $a\approx 253.27\unit{ms^{-2}}$. Furthermore, we cut out the poles of the sphere by limiting the colatitude angle $\vartheta$ to $\delta\leq\vartheta\leq\pi-\delta$ with $\delta\approx 0.01$.
To obtain a minimum energy configuration, we add an artificial linear friction term with frictional constant $\eta=50$, see \ref{appsec:partFriction}. As we will see, this is necessary because the energy loss due to electromagnetic radiation, which can be estimated by the Larmor formula
\begin{equation}
  P = \frac{e^2}{6\pi\epsilon_0c^3}a^2 \approx 5.708\cdot 10^{-54}\unit{\frac{Js}{m^3}}\cdot a^2,
  \label{eq:larmor}
\end{equation}
where we could use the centripetal acceleration $a=v^2/r$ for circular motion, is too small.

\begin{figure}[ht]
  \includegraphics[scale=0.15]{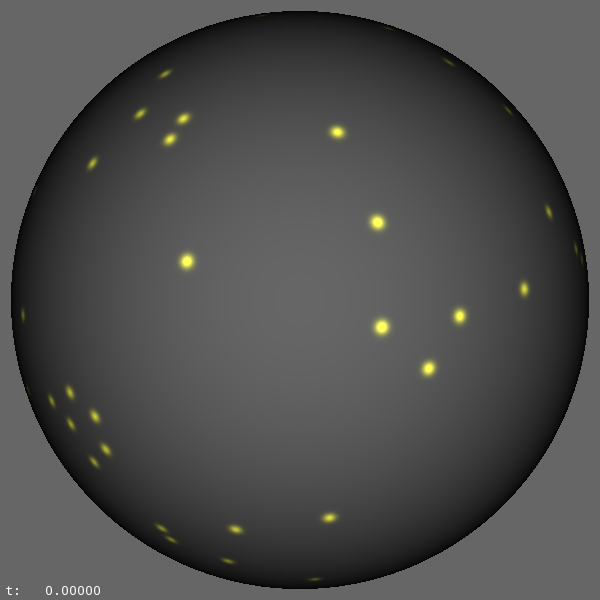}
  \includegraphics[scale=0.15]{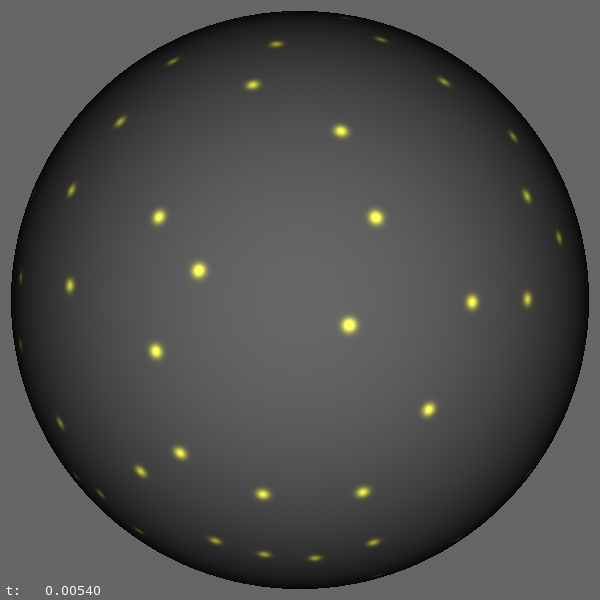}
  \includegraphics[scale=0.15]{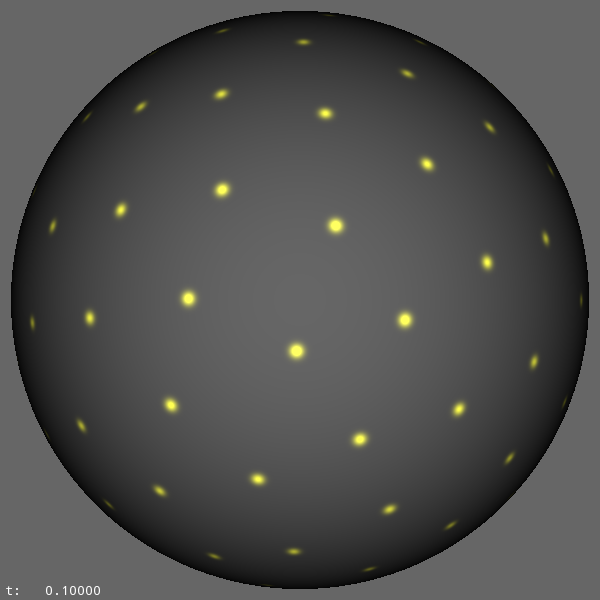}
  \includegraphics[scale=0.15]{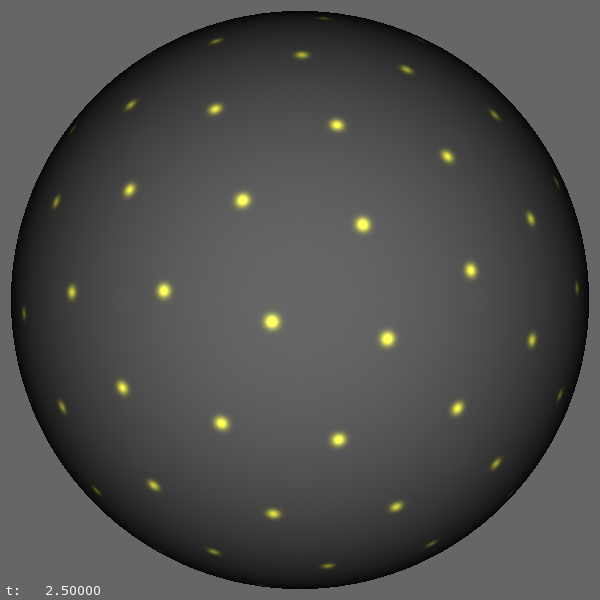}
  \caption{$N=128$ charged particles on a sphere at simulation times $t=\{0,0.0054,0.1,2.5\}\unit{s}$ with step size $\Delta t=10^{-4}\unit{s}$ and velocity-dependent friction $\eta=50$ (see \ref{appsec:partFriction}).}
  \label{fig:expSphere}
\end{figure}

At the beginning of the simulation, the particles are randomly distributed and have zero initial velocity, see figure~\ref{fig:expSphere}. The field energy $W|_{t=0}/(\kappa m_e)\approx 8.634620\cdot 10^3~\unit{m^{-1}}$, see equation \eref{eq:fieldEnergy}, and the kinetic energy $T/m_e=0$. 
When the simulation starts, the non-uniform distribution of the particles let them accelerate due to their mutual Coulomb interaction, and they reach a maximum kinetic energy of $T/m_e\approx 2.8467\cdot 10^5\unit{m^2s^{-2}}$ at $t\approx 0.0054\unit{s}$. Then, the mean velocity of a single particle is $v\approx 47.2\unit{ms^{-1}}$ and, thus, energy loss due to electromagnetic radiation and acceleration due to equation \eref{eq:lorentz} can be neglected.
At simulation time $t=2.5\unit{s}$, the field energy has dropped to $W/(\kappa m_e)\approx 7.393081\cdot 10^3\unit{m^{-1}}$ and the kinetic energy has reduced to $T/m_e\approx 2.433780\cdot 10^{-2}\unit{m^2s^{-2}}$, which is nearly impossible to observe visually, however.
After about $6.9$ seconds, the kinetic energy has dropped below $T/m_e=10^{-15}\unit{m^2s^{-2}}$.

In table~\ref{tab:E_sphere} we compare our minimum energy results for several number of particles $N$ with the literature values. Glasser and Every's~\cite{Glasser1992} estimation formula
\begin{equation}
  E(N) = \frac{N^2}{2}\left(1-aN^{-1/2}+bN^{-3/2}\right)
  \label{eq:eEstimate}
\end{equation}
with parameters $a=1.10461$ and $b=0.137$ from \cite{Morris1996} already gives a good approximation. The lowest energies for $110<N\leq 200$ from Morris et al.~\cite{Morris1996} are determined using a genetic algorithm.
\begin{table}[ht]
   \setlength{\tabcolsep}{0.35cm}
   \begin{tabular}{cccc} 
     \hline\\[-1.1em]
     \hline\\[-1em]
     $N$ & $W/(\kappa m_e)$ & $W_N$ & $E(N)$\\ \hline\\[-0.8em]
     $112$ & $5.618044887\cdot 10^3$ & $5.618044882\cdot 10^3$ & $5.618079704\cdot 10^3$\\
     $128$ & $7.393007443\cdot 10^3$ & $7.393007443\cdot 10^3$ & $7.392951914\cdot 10^3$\\
     $161$ & $1.183308476\cdot 10^4$ & $1.183308474\cdot 10^4$ & $1.183308683\cdot 10^4$\\
     $200$ & $1.843885657\cdot 10^4$ & $1.843884272\cdot 10^4$ & $1.843881429\cdot 10^4$
   \end{tabular}
   \caption{Minimum field energies $W/(\kappa m_e)$ for $N$ particles with charge $q=e$, time step $\Delta t=10^{-4}\unit{s}$ and frictional constant $\eta=50$; compared with values $W_N$ taken from Morris et al.~\cite{Morris1996} and estimation given by \eref{eq:eEstimate}. The latter two values were calculated using $\kappa m_e=1$.}
   \label{tab:E_sphere}
\end{table}

Depending on the initial random configuration and on the integration time step, the simulation does not always reach the same minimum energy configuration. In that case, the particles have to be either randomly distributed again or they have to be given a small jerk by pressing a key. In the just discussed example, our worst minimum energy for $N=128$  was about $W/(\kappa m_e)\approx 7.393166 \cdot 10^3\unit{m^{-1}}$.

\subsection{Particles on a torus}
\begin{figure}[ht]
  \includegraphics[scale=0.2]{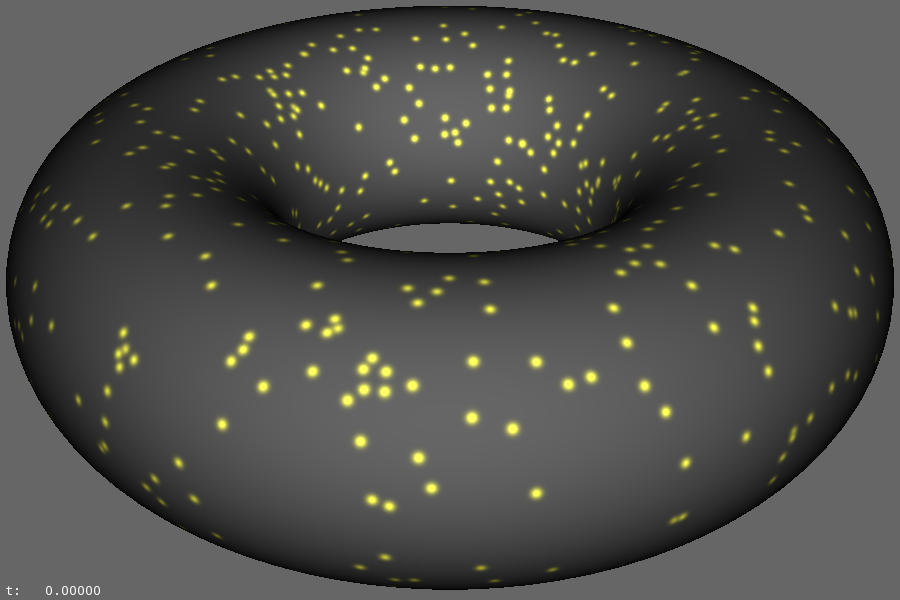}
  \includegraphics[scale=0.2]{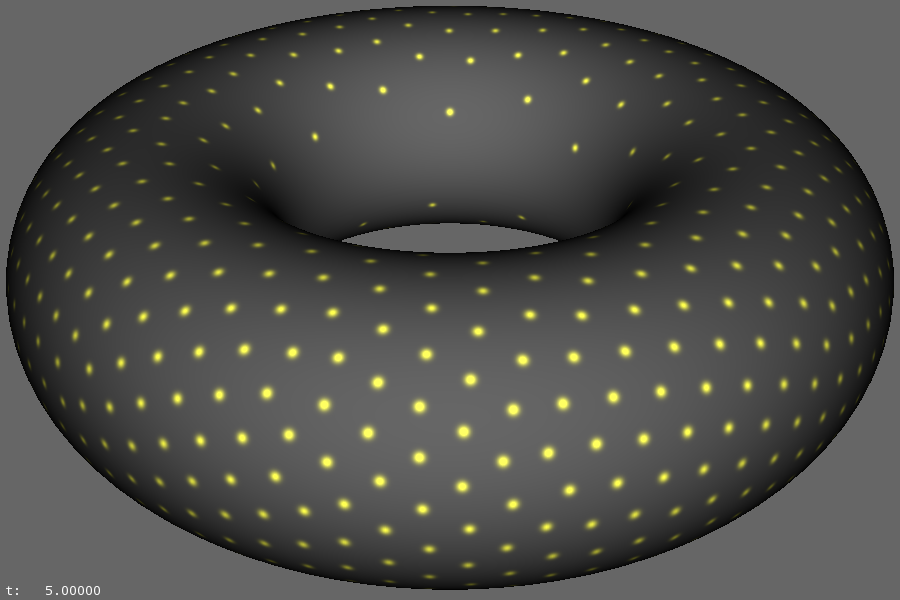}
  \caption{$N=1024$ charged particles on a torus at simulation times $t=\{0,5\}\unit{s}$ with step size $\Delta t=10^{-3}\unit{s}$ and velocity-dependent friction $\eta=200$.}
  \label{fig:expTorus}
\end{figure}

In the second example, we consider $N=1024$ particles $(q=e)$ on a torus with radii $R=2\unit{m}$ and $r=0.9\unit{m}$. As before, the particles are randomly distributed at the beginning of the simulation, see figure~\ref{fig:expTorus}. At $t\approx 30\unit{s}$, the kinetic energy has dropped below $T/m_e=2.134\cdot 10^{-4}\unit{m^2s^{-2}}$ and the field energy reads $W/(\kappa m_e)\approx 2.1760534\cdot 10^5\unit{m^{-1}}$.
Expectedly, all particles move to the outer side of the torus. Again, to stress the validity of our code, we compare our minimum energy configurations with the literature, see table~\ref{tab:E_torus}. As can be seen, we are in good agreement with the literature values and sometimes we have found even lower energies.
\begin{table}[ht]
   \setlength{\tabcolsep}{0.35cm}
   \begin{tabular}{cccc} 
     \hline\\[-1.1em]
     \hline\\[-1em]
     $N$ & $a=R/r$ & $W/(\kappa m_e)$ & $W_N$ \\ \hline\\[-0.8em]
      $20$ & $1.414$ & $1.029846718\cdot 10^2$ & $1.029846689\cdot 10^2$\\
      $20$ & $1.618$ & $1.098582569\cdot 10^2$ & $1.098582529\cdot 10^2$\\
     $100$ & $1.414$ & $3.082218463\cdot 10^3$ & $3.082217005\cdot 10^3$\\
     $100$ & $1.618$ & $3.295918820\cdot 10^3$ & $3.296043624\cdot 10^3$\\
     $415$ & $1.414$ & $5.660043959\cdot 10^4$ & $5.660070457\cdot 10^4$
   \end{tabular}
   \caption{Minimum field energies $W/(\kappa m_e)$ for $N$ particles with charge $q=e$, time step $\Delta t=0.0002\unit{s}$, and frictional constant $\eta=50$ on a torus with aspect $a=R/r$ and radius $R=1\unit{m}$; compared with values $W_N$ taken from \cite{thomsonApplet}.}
   \label{tab:E_torus}
\end{table}

\begin{figure}[ht]
  \includegraphics[scale=0.62]{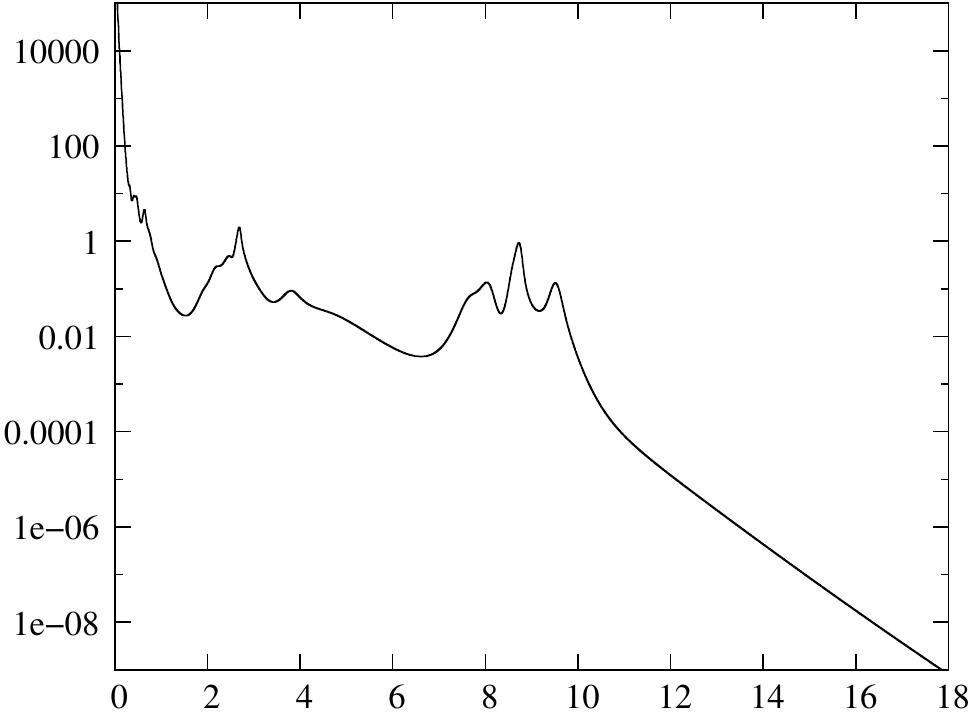}\quad
  \includegraphics[scale=0.62]{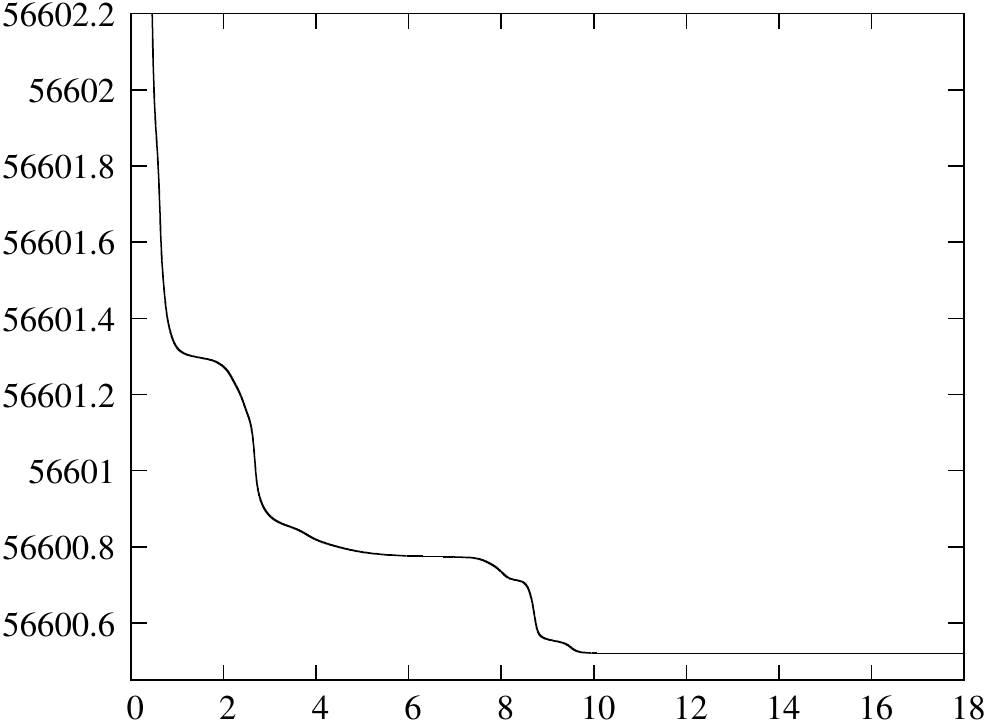}
  \caption{Kinetic energy $T/m_e$ in $\unit{m^2s^{-2}}$ (left) and field energy $W/(\kappa m_e)$ in $\unit{m^{-1}}$ (right) depending on time in $\unit{s}$ for the torus simulation with $N=415$ particles, aspect $a=1.414$, step size $\Delta t=0.01\unit{s}$, and a frictional constant $\eta=50$. The loss of kinetic energy is also due to the dissipative RK2 method.}
  \label{fig:torusEnergies}
\end{figure}

The dynamic evolution of the $N=415$ example of table~\ref{tab:E_torus} can be read from figure~\ref{fig:torusEnergies}. At the beginning of the simulation, there is a very short peak of high kinetic energy due to the particles that are accelerated from rest. After less than $0.8$ seconds, the kinetic energy has dropped below $T/m_e=1\unit{m^2s^{-2}}$. Then, at least from the visual impression, the particles do not move any more. However, there is still kinetic energy in the system which does not dissipate uniformly. Only after about $10$ seconds, the particles slowly settle down and reach a minimum energy configuration (plateau in figure~\ref{fig:torusEnergies}, right). 

\subsection{Two intertwined torii}
\begin{figure}[ht]
  \includegraphics[scale=0.237]{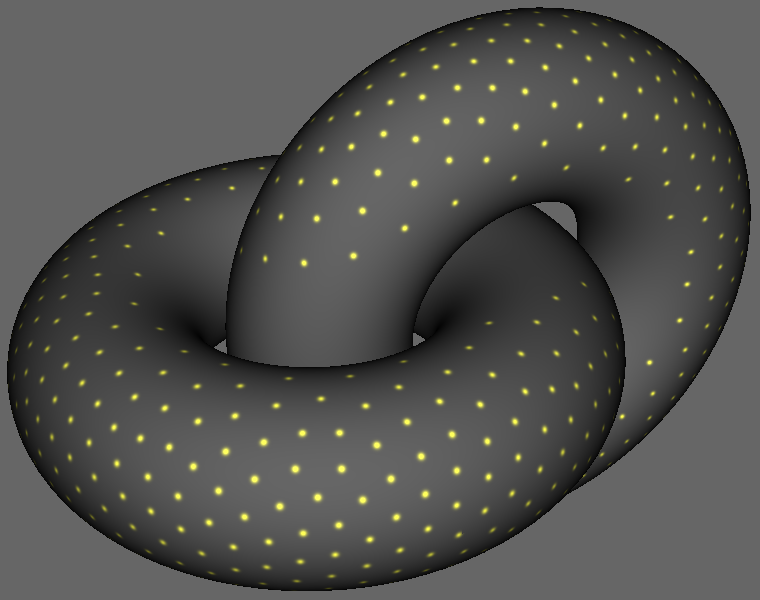}
  \includegraphics[scale=0.237]{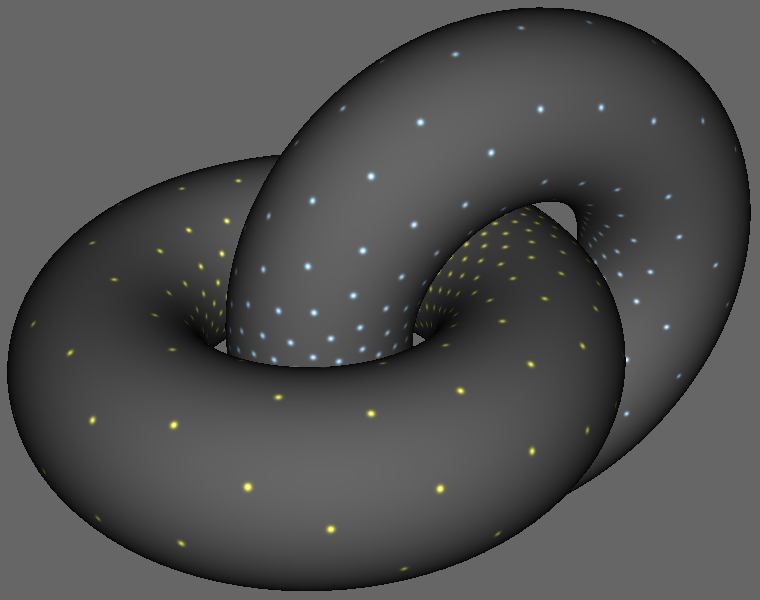}
  \caption{$N=2\times 512$ particles $(q=\pm e)$ on two intertwined torii with radii $R=2\unit{m}$ and $r=0.9\unit{m}$ of either the same (left) or opposite (right) charge.}
  \label{fig:expTorii}
\end{figure}

\begin{figure}[ht]
  \includegraphics[scale=0.237]{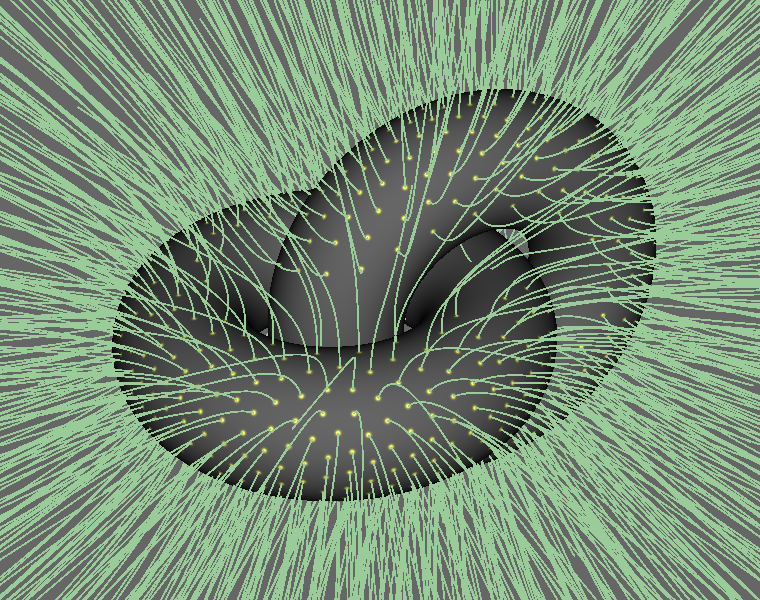}
  \includegraphics[scale=0.237]{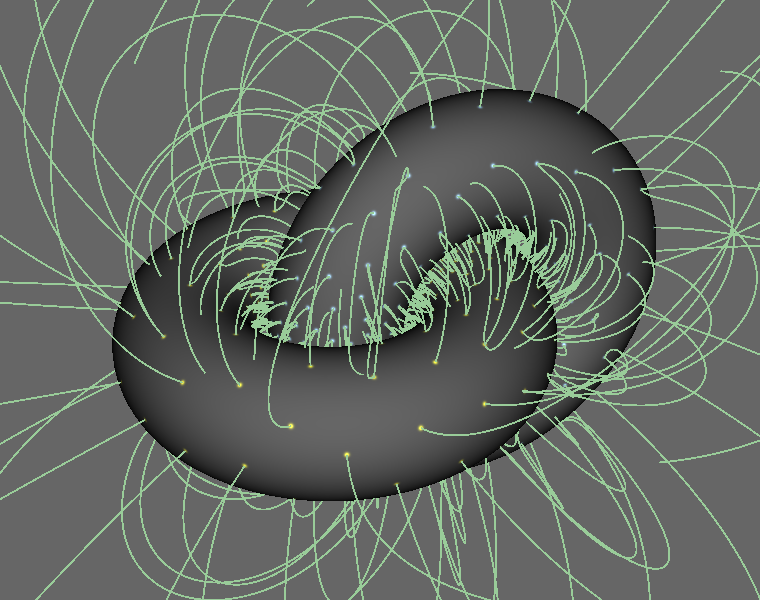}
  \caption{The same situation as in figure~\ref{fig:expTorii} but now with field lines (green) that were started close to the particle positions and perpendicular to the surface.}
  \label{fig:expToriiFL}
\end{figure}

A more intricate example is shown in figure~\ref{fig:expTorii} where two torii are intertwined and the particles have either the same or opposite charges. The corresponding field lines are shown in figure~\ref{fig:expToriiFL}. They were started close to the charged particles in the direction of the outer surface normal and were integrated with a constant step size along the total electric field of all particles. Please note that the field lines end after $n=800$ steps by default. And, as the object surface is no equipotential surface, field lines could also penetrate the object surface non-orthogonally.

% -----------------------------------------------------------------
\section{Exercises}\label{sec:exercises}
In the following we give a few suggestions for possible exercises that could be done using the ChaPaCS source code.

% ---------------------
\begin{cpExercise}
  Consider the Thomson problem situation where $N=128$ charged particles $(q=e)$ are located on a sphere. How does the minimum field energy vary with the size of the radius?
\end{cpExercise}

\begin{cpResult}
  When the charged particles are located in a minimum energy configuration, their relative positions on the sphere will not change when the radius of the sphere is changed. As the distance $d$ between two points on a sphere of radius $r$ is $d=2r\sin(\alpha/2)$, where $\alpha$ is the angular separation between both points, the field energy scales inversely proportional with the radius of the sphere: $W_{r_1}/W_{r_2}=r_2/r_1$. To test this with ChaPaCS, start with a minimum field energy and store the positions to disk. Restart the program with an other radius and load the previously stored positions. 
\end{cpResult}

% ---------------------
\begin{cpExercise}
  Adapt the {\tt set\_Particles()} method of the single sphere example such that one half of the particles have $q=e$ and the other ones have $q=10e$. To circumvent numerical instabilities, you have to reduce the time step to $\Delta t=2\cdot 10^{-5}\unit{s}$ and you have to increase the frictional constant to $\eta=500$. What happens?
\end{cpExercise}

\begin{cpResult}
  Because of the partially higher charges, the field energy rises considerably from $W_1/(\kappa m_e)\approx 7.3930\cdot 10^3\unit{m^{-1}}$ to $W_2/(\kappa m_e)\approx 2.1709\cdot 10^5\unit{m^{-1}}$. Additionally, the symmetric arrangement of the particles breaks down.
\end{cpResult}

% ---------------------
\begin{cpExercise}
   Determine the influence of the aspect $a=R/r$ in the torus example with $N=415$ particles on the minimum field energy.
\end{cpExercise}

\begin{cpResult}
   Unfortunately, the change of the minimum field energy cannot be calculated as easily as in the first exercise. When the aspect ratio is increased, the 'thickness' radius decreases and the particles move together which results in a higher field energy.\\[0.3em]
   \begin{tabular}{cccc} 
     \hline\\[-1.1em]
     \hline\\[-1em]
      $a=R/r$ & $W/(\kappa m_e)$  \\ \hline\\[-0.8em]
      $1.45$ & $5.73353\cdot 10^4$\\
      $1.5$  & $5.83289\cdot 10^4$\\
      $1.7$  & $6.20165\cdot 10^4$\\
      $2.0$  & $6.68266\cdot 10^4$\\
      $2.5$  & $7.34052\cdot 10^4$\\
      $3.0$  & $7.87078\cdot 10^4$\\
      $5.0$  & $9.29907\cdot 10^4$
   \end{tabular}
\end{cpResult}

% ---------------------
\begin{cpExercise}
   Determine a minimum energy configuration for $N=128$ charged particles $(q=e)$ on the sphere and save it to disk. Restart the program with this configuration. Set the frictional constant to zero and add an external electric field, $\tilde{\mathbf{E}}_y=100\unit{ms^{-2}}$. Slightly increase the electric field strength and discuss what happens.
\end{cpExercise}
\pagebreak

\begin{cpResult}
  At the beginning of the simulation, the particles are at rest. With the electric field turned on, the particles will be accelerated in the direction of the field. But because of the spherical constrained, the particle move together which increases the electric repulsion. Hence, the particles begin to oscillate.
  Due to numerical dissipation, the oscillation is damped and the particles come to rest after some time.
  If the field strength is increased, the oscillation amplitude becomes higher. Since the particles are constrained to the sphere and cannot linearly follow the field lines, the system becomes more and more chaotic. And since the step size is fixed, the numerical integration becomes unstable.
\end{cpResult}

\noindent {\bf Further exercises:}
\begin{itemize}
  \item Find the parametrization of an ellipsoid and determine the metric coefficients $g_{ij}$ and the Christoffel symbols $\Gamma_{ij}^k$. Expand the code to handle ellipsoids and find the minimum energy configurations.
  \item Construct a new scene with a small sphere hovering above a plane. 
  \item Construct a new scene with four small spheres at the corners of a quad (quadrupole). Study the distribution of particles on the spheres and the overall field lines.  
\end{itemize}

\section{Summary}
In this work we have developed the equations of motion of $N$ charged particles that are constrained to a curved two-dimensional surface but interact in three dimensions as usual. We have also given a short introduction how to implement the particle simulation using the compute capability of todays graphics boards. The source code (C/C++/CUDA) of the prototype implementation \emph{ChaPaCS} is freely available and can be easily extended by other curved surfaces.

\ack
This work was partially funded by Deutsche Forschungsgemeinschaft (DFG) as part of the Collaborative Research Centre SFB 716.

% -----------------------------------------------------------------
%   appendix
% -----------------------------------------------------------------
\appendix
% -----------------------------------------------------------------
\section{Units}\label{appsec:units}
For numerical computations we should know which order of magnitudes we have to deal with. If we use the electron mass $m_e$ and the electron charge $e$ as basis units, we could replace the mass and charge of a particle by dimensionless factors $m_A$ and $q_A$ as follows: $M_A=m_Am_e$ and $Q_A=q_Ae$.
For the equations of motion \eref{eq:eomAll}, we set
\begin{equation}
  \kappa := \frac{e^2}{4\pi\epsilon_0m_e} = \frac{e^2\mu_0c^2}{4\pi m_e}\approx 253.27\unit{\frac{{m}^3}{{s}^2}},
\end{equation}
where $m_e\approx 9.109\cdot 10^{-31}\unit{kg}$, $e\approx 1.6022\cdot 10^{-19}\unit{C}$, $\mu_0=4\pi 10^{-7}\unit{Vs/(Am)}$, and $c=299792458\unit{m/s}$. Furthermore, we have $\kappa m_e\approx 1.440\cdot 10^{-9}\unit{eV}\cdot\unit{m}= 1.440\unit{neV}\cdot\unit{m}$, and we combine the electric charge $e$ and the electron mass $m_e$ with the electric and magnetic fields:
\begin{equation}
  \tilde{\mathbf{E}}:=\frac{e}{m_e}\mathbf{E},\quad \tilde{\mathbf{B}}:=\frac{e}{m_e}\mathbf{B},
\end{equation}
with dimensions $[\tilde{\mathbf{E}}]=\unit{C/kg}\cdot\unit{N/C} = \unit{m/s^2}$ and $[\tilde{\mathbf{B}}]=\unit{C/kg}\cdot\unit{N/(Am)} = \unit{s^{-1}}$.

% -----------------------------------------------------------------
\section{Particle motion with friction}\label{appsec:partFriction}
Let the particle motion be damped by a velocity-dependent friction $\mathbf{F}_R=-h(v)\frac{\mathbf{v}}{v}$ with the frictional function $h$ depending on the value of the velocity $v=\|\mathbf{v}\|$. Then, the generalized friction $R_k$ reads
\begin{equation}
 R_k = -h(v)\frac{\mathbf{v}}{v}\cdot\frac{\partial\mathbf{r}}{\partial u_k}=-h(v)\frac{\mathbf{v}}{v}\cdot\frac{\partial\mathbf{v}}{\partial \dot{u}_k} = -\frac{h(v)}{v}\left<\dot{\mathbf{f}},\pdiff{\dot{\mathbf{f}}}{\dot{u}^k}\right>.
\end{equation}
For the linear friction $h(v)=\eta v$ with frictional constant $\eta$ and the time derivative of $\mathbf{f}$, compare \eref{eq:timeDiffF}, we obtain
\begin{equation}
 R_k = -\eta\sum_{j=1}^2\left<\pdiff{\mathbf{f}}{u^j},\pdiff{\mathbf{f}}{u^k}\right>\dot{u}^j=-\eta\sum_{j=1}^2g_{jk}\dot{u}^j.
\end{equation}
The Euler-Lagrange equation \eref{eq:eulerLagrange} now reads
\begin{equation}
 0 = \frac{d}{dt}\pdiff{L}{\dot{u}_A^i}-\pdiff{L}{u_A^i}-R_i.
\end{equation}
The right-hand side of the equation of motion \eref{eq:eomAll} must be extended by $-\eta\dot{u}^i$. As the parameters $u^i$ are dimensionless, $\eta$ is also dimensionless.
%The unit of $\eta$ depends on the unit of the coordinates $u^i$. As these could be different, we should use two separate frictional constants. But here, we use the same numerical value and use the respective unit if necessary.

% -----------------------------------------------------------------
%    
% -----------------------------------------------------------------
\section{Surface examples}\label{appsec:surfExp}
The following surface parametrizations are given in standard form, which means that the components of $\mathbf{f}$ are with respect to the global Cartesian coordinate system of $\mathbb{E}^3$. Thus, $\mathbf{f}=f^1\mathbf{e}_1+f^2\mathbf{e}_2+f^3\mathbf{e}_3$ with
\begin{eqnarray}
  \mathbf{e}_1=(1,0,0)^T,\quad \mathbf{e}_2=(0,1,0)^T,\quad \mathbf{e}_3=(0,0,1)^T,
\end{eqnarray}
and the center of the object equals the origin of the global coordinate system. However, for more elaborate scenes, the basis $\{\mathbf{e}_1,\mathbf{e}_2,\mathbf{e}_3\}$ can be oriented and translated arbitrarily.

% ------------------------------
% 
% ------------------------------
\subsection{Sphere}\label{appsec:sphere}
The surface of a sphere with radius $r$ can be parametrized using spherical coordinates $(u^1=\varphi,u^2=\vartheta)$:
\begin{equation}
 \mathbf{f}(\varphi,\vartheta)=r\left(\begin{array}{c}\sin\vartheta\cos\varphi\\ \sin\vartheta\sin\varphi\\ \cos\vartheta\end{array}\right)
\end{equation}
with $\varphi\in[0,2\pi)$, $\vartheta\in(0,\pi)$.
The metric coefficients $g_{ij}$ follow from \eref{eq:metricCoeffs},
\begin{equation}
 g_{11}=r^2\sin^2\vartheta,\quad g_{12}=0,\quad g_{22}=r^2,
\end{equation}
and the only non-vanishing Christoffel symbols of the second kind are
\begin{equation}
 \Gamma_{12}^1=\cot\vartheta,\quad \Gamma_{11}^2 = -\sin\vartheta\cos\vartheta.
\end{equation}

% ------------------------------
% 
% ------------------------------
\subsection{Torus}
A torus is defined by two radii, where $R$ is the radius of the main circle and $r$ is the 'thickness' radius,
\begin{equation}
 \mathbf{f}(\vartheta,\varphi)=\vecThree{(R+r\cos\vartheta)\cos\varphi}{(R+r\cos\vartheta)\sin\varphi}{r\sin\vartheta}.
\end{equation}
Here, $u^1=\vartheta\in[0,2\pi)$ and $u^2=\varphi\in[0,2\pi)$.
The metric coefficients are straightforward
\begin{equation}
  g_{11} = r^2,\quad g_{12} = 0,\quad g_{22} = (R+r\cos\vartheta)^2,
\end{equation}
and the non-vanishing Christoffel symbols of the second kind read
\begin{equation}
  \Gamma_{12}^2 = -\frac{r\sin\vartheta}{R+r\cos\vartheta},\quad \Gamma_{22}^1 = \frac{(R+r\cos\vartheta)\sin\vartheta}{r}.
\end{equation}

 -----------------------------------------------------------------
%                           thebibliography
% -----------------------------------------------------------------
\section*{References}
\bibliographystyle{unsrt}
\bibliography{lit_cp}
\end{document}